\documentclass[conference]{IEEEtran}
\IEEEoverridecommandlockouts

\usepackage{cite}
\usepackage{amsmath,amssymb,amsfonts}
\usepackage{algorithmic}
\usepackage{graphicx}
\usepackage{textcomp}
\usepackage{xcolor}
\usepackage{multirow}
\usepackage{tabularx}
\usepackage{url}
\usepackage{svg}
\usepackage[utf8]{inputenc}
\usepackage{graphicx} 
\usepackage{tikz}     
\usepackage{booktabs} 
\usepackage{graphicx} 
\def\BibTeX{{\rm B\kern-.05em{\sc i\kern-.025em b}\kern-.08em
    T\kern-.1667em\lower.7ex\hbox{E}\kern-.125emX}}
\begin{document}

\title{Open Source State-Of-the-Art Solution for Romanian Speech Recognition
}

\author{\IEEEauthorblockN{Gabriel Pîrlogeanu, Alexandru-Lucian Georgescu, Horia Cucu}
\IEEEauthorblockA{\textit{Speech and Dialogue Research Laboratory, POLITEHNICA Bucharest} \\
Bucharest, Romania \\
\{gabriel.pirlogeanu, lucian.georgescu, horia.cucu\}@upb.ro}
}
\maketitle

\begin{abstract}
In this work, we present a new state-of-the-art Romanian Automatic Speech Recognition (ASR) system based on NVIDIA’s FastConformer architecture—explored here for the first time in the context of Romanian. We train our model on a large corpus of, mostly, weakly supervised transcriptions, totaling over 2,600 hours of speech. Leveraging a hybrid decoder with both Connectionist Temporal Classification (CTC) and Token-Duration Transducer (TDT) branches, we evaluate a range of decoding strategies including greedy, ALSD, and CTC beam search with a 6-gram token-level language model. Our system achieves state-of-the-art performance across all Romanian evaluation benchmarks, including read, spontaneous, and domain-specific speech, with up to 27\% relative WER reduction compared to previous best-performing systems. In addition to improved transcription accuracy, our approach demonstrates practical decoding efficiency, making it suitable for both research and deployment in low-latency ASR applications.
\end{abstract}


\section{Introduction}

Automatic Speech Recognition (ASR) has undergone a paradigm shift over the past decade, driven by the rise of end-to-end architectures and the increasing availability of large-scale datasets. Models such as RNN-Transducer, Transformer-Transducer, wav2vec, Whisper, Conformer~\cite{prabhavalkar2023endtoendspeechrecognitionsurvey} have dramatically improved recognition accuracy across many languages. Most recently, Speech Large Language Models (SpeechLLMs) \cite{peng2025surveyspeechlargelanguage} have further advanced the field by integrating multimodal and multilingual supervision at unprecedented scale.

Besides architectural innovations, decoding strategies have also evolved significantly. Beyond the ubiquitous Connectionist Temporal Classification (CTC) and RNN-T approaches, recent work has demonstrated the utility of advanced techniques such as Alignment-Length Synchronous Decoding (ALSD)~\cite{alsd}, token-duration modeling~\cite{tdt} and hybrid decoding frameworks that combine the strengths of multiple objectives~\cite{hybrid}. Furthermore, the integration of external language models (LMs), particularly n-gram or neural LMs, has proven essential in bridging acoustic and linguistic gaps, especially for under-resourced languages.

Despite these advances, Romanian remains a low-resource language in the context of ASR. Earlier efforts have primarily focused on hybrid HMM-DNN systems, which established strong baselines on several benchmarks~\cite{reterom}. While neural ASR systems have recently been applied to Romanian, they often rely on architectures or training strategies that do not reflect the latest developments in the field. For instance, DeepSpeech~\cite{deepspeech-ro}, wav2vec-based approaches~\cite{wav2vec-ro} and Whisper adaptations~\cite{whisper-ro} have shown promising results, but no prior work has explored the Conformer~\cite{conformer} or  FastConformer~\cite{fastconformer} architecture for Romanian speech, nor has there been an exhaustive exploration of decoding strategies tailored to this language.

The scarcity of manually annotated Romanian data poses a significant barrier to fully supervised learning. The largest publicly available dataset, the Read Speech Corpus (RSC)~\cite{georgescu-etal-2020-rsc}, provides high-quality transcriptions but remains modest in size. However, recent efforts have expanded coverage across domains, dialects, and speech styles. Resources such as CoBiLiRo~\cite{cristea-etal-2020-cobiliro}, CoRoLa~\cite{mititelu-etal-2014-corola}, and USPDATRO~\cite{uspdatro} have introduced more spontaneous and dialectal content. Notably, Georgescu et al. \cite{reterom} demonstrated that training on over 600 hours of mostly weakly labeled read and spontaneous speech can significantly enhance ASR robustness and generalization. However, the authors also observed a degradation in spontaneous speech recognition performance when adding $2000$ hours of oratory speech.

Large-scale weak supervision, such as learning from pseudo-labels or partially aligned transcripts, has emerged as a powerful strategy for under-resourced languages~\cite{synnaeve2020, radford2022-whisper,wav2vec}. These techniques enable the use of vast audio corpora with minimal human supervision, thereby bridging the gap between resource-rich and resource-poor settings. Whisper~\cite{radford2022-whisper}, for instance, exemplifies how weakly supervised multilingual training can yield high-quality models even with noisy labels.

In this paper, we propose the first adaptation of NVIDIA’s FastConformer architecture for Romanian ASR. We fine-tune a 110M parameter hybrid CTC-TDT~\cite{fastconformer,hybrid} model using over 2600 hours of Romanian speech, composed of both high-quality manual transcriptions and weakly labeled data obtained through partial alignment techniques. Our study not only benchmarks transcription accuracy through Word Error Rate (WER), but also evaluates computational efficiency via the Real-Time Factor (RTFx). We explore a spectrum of decoding strategies—including CTC greedy, TDT greedy, TDT with ALSD, and CTC beam search with an external 6-gram language model—leveraging the decoder's hybrid nature to gain insight into performance trade-offs.

Our system achieves state-of-the-art results across seven diverse Romanian ASR benchmarks, covering read, spontaneous, oratory, and underrepresented speech. These results highlight the effectiveness of the FastConformer architecture when combined with scalable training and decoding pipelines, offering a powerful new baseline for Romanian speech recognition research. To promote continued progress in Romanian speech processing, we will publicly release our trained model, along with comprehensive training and inference recipes, and the standardized evaluation datasets\footnote{\url{https://github.com/gabitza-tech/SpeD-RoASR}}.

\section{Methodology}

\subsection{Encoder Architecture}
\label{sec:encoder-architecture}
The encoder used in this work is based on the FastConformer~\cite{fastconformer}, a highly efficient variant of the Conformer~\cite{conformer} architecture designed for ASR. The Conformer architecture itself extends the Transformer~\cite{transformer} by incorporating convolutional modules to better capture local dependencies in speech, which are often missed by purely self-attentive models. Each Conformer block consists of a feed-forward module, a multi-headed self-attention module with relative positional encoding, a convolution module, and a second feed-forward module, all connected via residual connections and layer normalization. This design enables modeling of both global and local temporal relationships, making the Conformer particularly effective for speech tasks.

Building on this foundation, the FastConformer introduces architectural optimizations aimed at reducing computational cost while preserving accuracy. The FastConformer encoder addresses these limitations by redesigning the downsampling schema and optimizing key architectural components to improve both training and inference efficiency. One of the primary modification of the FastConformer is the introduction of an eightfold downsampling step at the beginning of the encoder using a stack of three depthwise separable convolutional layers. This reduces the input sequence length significantly, thereby decreasing the computational burden on subsequent attention and convolution blocks without sacrificing model accuracy. Additionally, FastConformer reduces the convolutional kernel size from 31 to 9 and decreases the number of channels in the subsampling layers from 512 to 256. These changes lower the model’s parameter count and operation cost while preserving its representational capacity.

FastConformer enhances efficiency for long-form audio by replacing global attention with limited-context attention and a global token, inspired by the Longformer~\cite{longformer}. This enables processing of sequences up to 11 hours in a single forward pass while maintaining or improving word error rates. Importantly, the overall Conformer architecture and block design remain unchanged, allowing FastConformer to preserve the strong performance of its predecessor while delivering up to 2.8$\times$ faster inference with substantially lower compute demands.

In this work, we adopt a 17-layer FastConformer encoder, resulting in approximately 110 million parameters.

\subsection{Decoder}



\begin{figure}[t]
    \centering
    \includegraphics[width=0.5\textwidth]{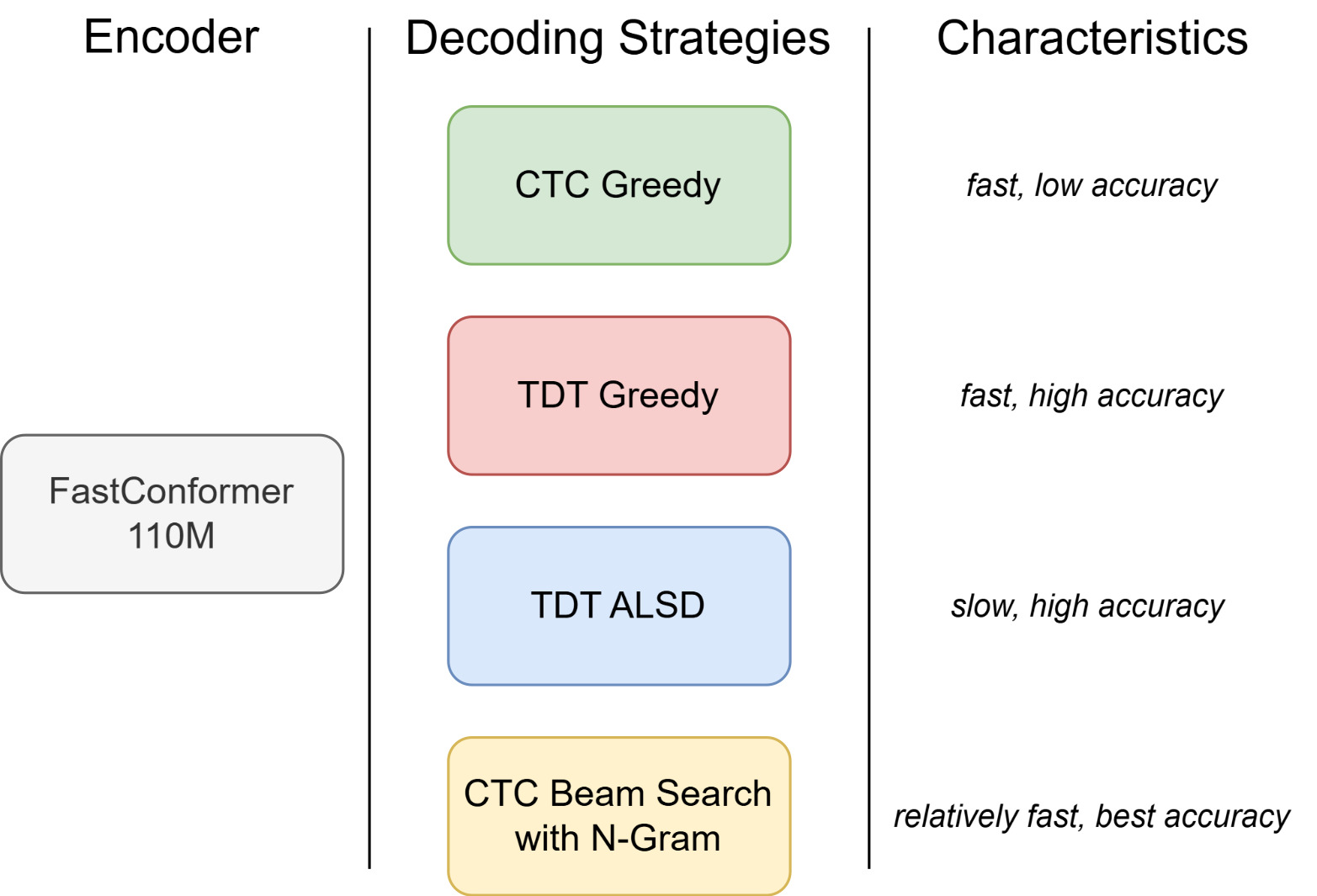} 
    \caption{Comparative analysis of the decoding strategies explored in this work, evaluated in terms of ASR accuracy and inference latency on Romanian speech.}
    \label{fig:decoding-strats} 
\end{figure}

In this section, we provide a detailed analysis of the decoder architectures employed in this study, along with advanced decoding strategies aimed at improving ASR performance. Figure~\ref{fig:decoding-strats} offers a comparative overview of the decoding approaches evaluated on Romanian speech, highlighting their respective trade-offs in terms of latency and recognition accuracy.

The Recurrent Neural Network Transducer (RNN-T)~\cite{rnnt} loss is foundational in end-to-end ASR, capable of jointly learning acoustic and language modeling. To address issues like alignment ambiguity and training instability, the Token-Duration Transducer (TDT)~\cite{tdt} explicitly models token durations, improving temporal alignment, convergence, and accuracy. While a greedy TDT decoder offers good computational speed, its sub-optimal search can yield significantly lower accuracy by easily getting stuck to local maxima. For efficient yet accurate inference, the Alignment-Length Synchronous Decoding (ALSD)~\cite{alsd} method for RNN-T/TDT models is preferred over greedy approaches. ALSD employs a beam search controlled by total alignment length, providing a superior balance between speed and performance. Its key advantages include significantly improved accuracy over greedy methods and enhanced computational efficiency compared to standard time-synchronous beam decoding, often with reduced tuning. However, ALSD remains computationally more intensive than pure greedy decoding, representing a trade-off for its higher fidelity results.

Connectionist Temporal Classification (CTC)~\cite{ctc} defines a common ASR loss function. While a greedy CTC decoder is fast, it yields sub-optimal transcripts as it performs a limited search. For improved accuracy, Beam Search CTC decoding explores multiple hypotheses. A key advantage is its integration with an external Language Model (LM), which re-scores hypotheses to enhance linguistic plausibility and achieve state-of-the-art performance. However, beam search incurs higher computational cost and latency than greedy decoding, and requires careful tuning of LM interpolation weights. Despite these drawbacks, the accuracy gains typically favor beam search with an external LM for robust ASR systems.

Following the approach proposed in~\cite{hybrid}, we employ a unified and efficient hybrid decoder architecture that integrates both a Connectionist Temporal Classification (CTC)~\cite{ctc} decoder and a Token-Duration Transducer (TDT)~\cite{tdt} decoder, sharing a common encoder. This design enables flexible decoding at inference time, allowing for the selection of the most appropriate strategy depending on the target application. Beyond its versatility, this hybrid framework offers several practical advantages: it eliminates the need to train and maintain separate models, thereby reducing computational overhead; it accelerates the convergence of the CTC decoder; and it enhances the overall recognition accuracy of both decoding branches, likely due to the benefits of joint optimization. During training, the final loss is computed as a weighted sum of the individual losses from the CTC and TDT decoders, encouraging the model to learn representations that are beneficial to both objectives.

\section{Experimental Setup}

In this section we will offer a comprehensive description of the datasets employed in this study, the training strategy, the language modeling step, the baseline systems, and we also describe the evaluation protocol.

\subsection{Speech Datasets}
\label{sec:datasets}

\begin{table}[ht]
\centering
\caption{The composition of the training, validation and evaluation datasets. For each subset, we report the total duration in hours, as well as the average utterance duration in seconds. }
\label{tab:dataset_summary}
\begin{tabular}{llcc}
\toprule
\textbf{Subset} & \textbf{Datasets} & \textbf{Total Dur. [h]} & \textbf{Avg. Utt. Dur. [s]} \\
\midrule
\multirow{5}{*}{Training} 
    & \texttt{RSC-train} & 93.7 & 2.5 \\
    & \texttt{CoBiLiro-train} & 30.3 & 2.2 \\
    & \texttt{CoRoLa-train} & 56.3 & 7.7 \\
    & \texttt{SSC-train} & 407.3 & 2.7 \\
    & \texttt{CDEP-train} & 2048.5 & 4.4 \\

\midrule
\multirow{3}{*}{Validation} 
    & \texttt{SSC-dev} & 10.9 & 25.1 \\
    & \texttt{CoRoLa-dev} & 27.3 & 32.3 \\
\midrule
\multirow{7}{*}{Test} 
    & \texttt{RSC-eval} & 5.2 & 7.6 \\
    & \texttt{SSC-eval1} & 3.5 & 4.12 \\
    & \texttt{SSC-eval2} & 1.5 & 54.7 \\
    & \texttt{CDEP-eval} & 4.9 & 60 \\
    & \texttt{CV21-RO} & 4.6 & 4.2 \\
    & \texttt{Fleurs-RO} & 2.5 & 10.3 \\
    & \texttt{USPDATRO} & 4.3 & 5.9 \\
\bottomrule
\end{tabular}

\end{table}

In Table~\ref{tab:dataset_summary} we present all the datasets used in this study for training, validation and evaluation, comprising of both read and spontaneous speech, with annotations obtained either manually or automatically. 

The training dataset is comprised of subsets also explored in \cite{reterom}: the Read Speech Corpus (\texttt{RSC-train})~\cite{georgescu-etal-2020-rsc}, the Spontaneous Speech Corpus (\texttt{SSC-train}) - comprised of 4 subsets, the Chamber of DEPuties Corpus (\texttt{CDEP-train}), the Bimodal Corpus for Romanian Language (\texttt{CoBiLiRo-train})~\cite{cristea-etal-2020-cobiliro} and the Corpus of the Contemporary Romanian Language (\texttt{CoRoLa-train})~\cite{mititelu-etal-2014-corola}. Due to architectural limitations, we limit the durations of the training audio files to a minimum of $0.1$s and a maximum of $20$s. Notably, the majority of this corpus consists of automatically generated annotations, obtained by aligning two ASR systems over the \texttt{SSC-train} and \texttt{CDEP-train} datasets, amounting to approximately $2455$ hours of annotations.
 In total, the training set consisted of around $2636$h, with $2.4$M utterances and an average file duration of $3.9$s.

The validation set has an important role in both training monitoring, as well as subsequent hyper-parameter tuning for the decoding strategies. In order to better model real life distributions of offline speech recordings, we choose audio files that are longer than $20$s. We also want to focus on spontaneous speech in this study, therefore we select the recordings from the SSC (\texttt{SSC-dev}) and CoRoLa (\texttt{CoRoLa-dev}) datasets. Our development set totals around $38$h and an average duration fo $29.85$s.

We employ an exhaustive evaluation over multiple Romanian speech datasets, containing both read and spontaneous speech. For read speech, we evaluate on the test set of the RSC (\texttt{RSC-eval}) dataset and for oratory speech (formal public speaking--Chamber of Deputies speech), we utilize the CDEP (\texttt{cdep-eval}) dataset. For spontaneous speech, we utilize the \texttt{SSC-eval1} and \texttt{SSC-eval2}~\cite{reterom} evaluation sets, as well as the \texttt{USPDATRO} dataset~\cite{whisper-ro,uspdatro}. We also evaluate on the Romanian test subsets of two large multilingual datasets: Common Voice 21.0 Romanian (\texttt{CV21-RO})~\cite{ardila-etal-2020-common}  and Fleurs Romanian (\texttt{Fleurs-RO})\cite{fleurs}.

\subsection{ASR Model setup}
\label{sec:asr_model}

Several prior studies have demonstrated that initializing speech processing models—such as those used for Automatic Speech Recognition (ASR) or Speech Translation—from pre-trained models on high-resource languages (e.g., English) significantly improves performance on low-resource languages~\cite{pretrain-asr}. This transfer learning approach is particularly effective when leveraging models trained via Self-Supervised Learning (SSL) on large-scale English audio corpora. Such pre-trained encoders capture universal low-level acoustic representations (e.g., phonetic features) that generalize well across languages, thereby providing a strong foundation for fine-tuning on target languages with limited labeled data. In contrast, models trained from scratch on low-resource languages often struggle to learn such robust representations due to insufficient training data.

We initialize our model from the 110M variant of the Parakeet Hybrid TDT-CTC architecture from Nvidia's NeMo toolkit~\cite{nemo-toolkit}. This model was pretrained in a SSL manner on the Librilight dataset~\cite{librilight}, then finetuned for offline speech recognition on $36$k hours of English annotated recordings. It has a tokenizer of 1024 BPE tokens.

In order to fine-tune this model on Romanian, we built a new tokenizer using the $2.4$M annotations from the speech training sets, as well as $24.6$M texts from a cleaned version of the \texttt{news-corpus} used in~\cite{reterom}, which will be further discuss in Section~\ref{sec:language_modeling}. We clean the text corpora in order to keep only the 31 official characters of the Romanian alphabet, alongside the hyphen (``-``) character. We built a tokenizer with a vocabulary size of 1024 using the SentencePiece~\cite{sentencepiece} toolkit, limiting subwords to a maximum of 5 subword tokens.

During training, besides the $2636$ hours of training speech, we also add noise augmentations using a $6$h Freesound subset from the MUSAN dataset~\cite{musan}, with an augmentation probability of $0.2$ and SNR in the range of $10$ to $30$. Additionally, we add speed perturbations in the range $0.9$ to $1.1$, with a $0.4$ augmentation probability. We also perform spectogram augmentations using SpecAugment and SpecCutout~\cite{specaugment}. Due to the fact that the decoder architecture includes both a TDT and CTC decoder, we set the weight of the CTC loss to $0.3$, when computing the combined hybrid loss. For the training strategy, we utilize the weighted Adam (AdamW)~\cite{adamw} optimizer with an initial learning rate of $2.0$ and a weight decay of $10^{-3}$. We use a Noam Annealing scheduler, with $10$k warming steps.

We train the model for 30 epochs with a batch size of $32$ and a gradient accumulation factor of $8$. Training is performed on a $24$GB NVIDIA RTX $4090$ GPU using BFloat16 precision, resulting in an epoch duration of approximately $5.5$ hours. The final model is derived through checkpoint averaging over the 10 checkpoints that achieve the lowest validation WER.


\subsection{Language Modeling and Decoding}
\label{sec:language_modeling}

Language modeling using n-grams helps automatic speech recognition (ASR) by predicting the most likely sequence of words based on context, thereby improving accuracy in distinguishing between acoustically similar words. Therefore, we train a token n-gram model using the KenLM toolkit~\cite{kenlm}. The unigrams are based on the ASR model's tokenizer. Similar to the tokenizer building, we use $2.4$M lines from the training annotations, as well as a cleaned version of the \texttt{news\_corpus} used in~\cite{reterom} (formed from \texttt{news002} and \texttt{news2020}). The unprocessed corpus contained over $1.4$M words in its lexicon. In order to reduce the dataset's size and remove unnecessary words, we limit the lexicon to the most frequent $500$k words by appearance, resulting in a corpus of $24.6$M lines.

With the $27$M input lines, we train a 6-gram token LM model. The resulting n-gram reaches a disk size of approx. $15$GB in the binarized form, leading to a significant memory consumption. We decide to prune the 6-gram model by the following scheme: $[0,1,3,5]$. This means that we drop bigrams that appear only once, trigrams that appear 3 times or less and 4/5/6-grams that appear 5 times or less. Pruning the model leads to a reduction in memory footprint to $2$GB.

We utilize this 6-gram model in a CTC beam decoding strategy. We tune the decoding hyper-parameters on the $38$h validation subset. For the CTC-beam decoding strategy, we set the beam size to $32$, the language model weight $\alpha$ to $0.9$ and the sequence length penalty score $\beta$ to $2$. For the TDT-ALSD strategy, we do not utilize and external language model and we only tune the beam size to $32$.

\subsection{Baseline Systems}

To assess the effectiveness of our proposed method, we compare it against state-of-the-art ASR systems that have been evaluated on established Romanian speech recognition benchmarks. The first baseline is a hybrid HMM-DNN system implemented using the Kaldi toolkit~\cite{reterom,kaldi}, which features a 13-layer Time-Delay Neural Network (TDNN) as the acoustic model. Decoding is performed using a 3-gram language model, followed by rescoring with an RNN-based language model. The acoustic model is trained on over 600 hours of Romanian speech data drawn from the \texttt{RSC}, \texttt{SSC}, \texttt{CoBiLiRo}, and \texttt{CoRoLa} corpora, while the language models are trained on a corpus of approximately 600 million words. In~\cite{reterom}, the authors further investigate the impact of incorporating an additional 2000 hours of speech from the \texttt{CDEP} dataset; however, their findings indicate that this addition led to a degradation in transcription quality for spontaneous speech.

The second baseline leverages a Whisper-based architecture, specifically the Whisper-large-v2 model with 1.55 billion parameters, fine-tuned on Romanian data (denoted as ``RoWhisper-large-v2'')~\cite{whisper-ro}. This model was evaluated on several standard Romanian ASR test sets, as well as a newly introduced dataset targeting underrepresented Romanian dialectal and spontaneous speech, \texttt{USPDATRO}. For this baseline, we report results obtained using beam search decoding.

\subsection{Data Preprocessing and Evaluation Protocol}
\begin{table*}[t]
\centering
\caption{Comparison of ASR system performance on seven Romanian evaluation datasets. Word Error Rate (WER) is reported as a percentage, with lower values indicating better transcription accuracy. Real-Time Factor (RTFx) is also reported, where higher values correspond to faster inference speed. * indicates that we report the value for RoWhisper-medium~\cite{whisper-ro}.}
\renewcommand{\arraystretch}{1.3}
\resizebox{\textwidth}{!}{%
\begin{tabular}{cc|ccccccc|c}
 & &  
\multicolumn{7}{c|}{\textbf{Evaluation Datasets [WER]}} & \raisebox{-0.8\totalheight}{\textbf{RTFx}}  \\
\textbf{Architecture} & \textbf{Decoding Strategy} &   \textbf{RSC-eval} & \textbf{SSC-eval1} & \textbf{SSC-eval2} & \textbf{CDEP-eval} & \textbf{CV-21} & \textbf{Fleurs-RO}  & \textbf{USPDATRO} & 
\\
\hline
HMM-DNN (TDNN)~\cite{reterom} & N-gram + RNN rescoring   & \underline{1.90} & 9.40 & 11.40 & 5.40 & --  & -- & -- & -- \\
RoWhisper-large-v2~\cite{whisper-ro} &  Beam & 3.09 & 25.05 & 61.46  & 62.83 & 9.31  & -- & 28.00* & -- \\
\hline
\multirow{3}{*}{\raisebox{1.5\totalheight}{Parakeet Ro 110M TDT (ours)}} 
 & Greedy  & 2.16 & 9.08 & \underline{10.85} &  4.20 & 3.57 & 10.61 & \underline{24.08} & \underline{126.15} \\
 & ALSD  & 2.05  &  \underline{8.64} & 10.88 & \underline{4.17} & \underline{3.38}  & \underline{10.16} & 24.3 & 66.63\\
\hline
\multirow{4}{*}{\raisebox{2.5\totalheight}{Parakeet Ro 110M CTC (ours)}} 
 & Greedy & 2.57 & 10.10 &  12.65  & 4.80 & 4.20 & 11.85 & 27.80  & \textbf{130.55} \\
 & Beam Token N-gram & \textbf{1.73}  & \textbf{8.12}  & \textbf{10.75} & \textbf{3.92} & \textbf{3.29} & \textbf{8.85} & \textbf{23.4}  &  109.46 \\

\end{tabular}
}
\label{tab:asr_results}
\end{table*}




To ensure a consistent evaluation across all datasets, we perform text normalization on the n-gram language model corpus, as well as on both training and evaluation annotations. Specifically, we retain only the lowercase forms of the 31 official characters of the Romanian alphabet, along with the hyphen character. All other punctuation marks and special symbols are removed. Additionally, for datasets such as \texttt{Fleurs-RO} and \texttt{USPDATRO}, we apply numeral-to-text conversion in order to unify numeric expressions across all corpora. On the audio processing side, the model accepts $16k$Hz mono-channel audio (wav) files as input.

We evaluate automatic speech recognition performance using the Word Error Rate (WER), a widely adopted metric defined as the total number of substitutions, deletions, and insertions divided by the number of words in the reference transcript. WER provides an intuitive measure of transcription accuracy, where lower values indicate higher fidelity to the ground truth. Due to its simplicity and interpretability, WER remains a standard benchmark for comparing ASR models across different datasets.

In addition to accuracy, we assess the computational efficiency of ASR models using the Real-Time Factor (RTFx), which measures the speed of transcription relative to the duration of the audio. For example, an RTFx of $\times100$ indicates that the system processes audio 100 times faster than its actual length. Unlike WER, RTFx captures the practical runtime efficiency of a model and is crucial in deployment scenarios. All decoding strategies are evaluated under a common setup: a 24 cores Intel i9-13900KF CPU-only environment with a batch size of 64. The final RTFx values are computed as the average over approximately 13,000 audio files, spanning durations from 0.2 to 260.4 seconds, across seven evaluation datasets.

\section{Results}


We evaluate our proposed approach on seven Romanian speech recognition datasets, benchmarking it against two baseline systems: a hybrid HMM-DNN (TDNN) model and a fine-tuned Romanian \texttt{Whisper-large-v2} model. Table~\ref{tab:asr_results} reports the Word Error Rate (WER) for each evaluation dataset, along with the Real-Time Factor (RTFx) computed over the concatenated evaluation sets using a CPU-only configuration.




As anticipated, the most efficient configuration in terms of latency is the CTC greedy decoding strategy. However, this comes at the cost of recognition accuracy, as it yields the highest Word Error Rate (WER) among the evaluated setups. Despite the absence of any language modeling, this configuration performs comparably to more complex decoding strategies and substantially outperforms the fine-tuned \texttt{RoWhisper-large-v2} model across all evaluation datasets. In comparison with the Kaldi-based baseline, the CTC greedy decoder achieves a relative WER reduction of 12.2\% on the \texttt{CDEP-eval} dataset.

The TDT greedy decoding setup yields consistent improvements over the baseline systems across all evaluation sets, with the exception of the \texttt{RSC-eval} dataset, where the HMM-DNN model achieves a lower WER. This decoding strategy serves as an effective trade-off between the simplicity of the CTC greedy approach—which makes frame-level independent predictions—and the more computationally intensive CTC beam search with external language modeling. Utilizing an internal language model, the TDT greedy decoder provides competitive performance close to that of the best-performing setup, while maintaining faster inference speed and eliminating the need for external language model training or tuning. Notably, the TDT decoding strategy demonstrates the model's ability to effectively leverage the 2000 hours of oratory speech from the \texttt{CDEP-train} dataset. In contrast to the baseline HMM-DNN system\cite{reterom}—which exhibited degraded performance on spontaneous speech when incorporating this domain—the FastConformer acoustic model benefits from the inclusion of oratory data, leading to improved performance even on spontaneous speech.

Finally, when employing the TDT decoder in conjunction with the ALSD strategy, we observe modest improvements in WER, compared to the greedy version, on most evaluation datasets, accompanied by a significant increase in latency as reflected in the RTFx values. While this approach is tuning-free and offers marginal transcription quality gains, its elevated computational cost may limit its practical applicability in latency-sensitive scenarios.

Our best-performing configuration, which employs CTC beam search decoding with a 6-gram token-level language model, achieves state-of-the-art performance across all evaluation datasets. Specifically, we observe a relative WER reduction of 9\% on the read speech dataset (\texttt{RSC-eval}), and a 27\% relative improvement on the oratory speech dataset (\texttt{CDEP-eval}). Furthermore, we obtain consistent gains on spontaneous speech datasets, with relative improvements of 14\% and 6\% on \texttt{SSC-eval1} and \texttt{SSC-eval2}, respectively. For the Romanian subset of multilingual corpora, our system achieves 3.3\% WER on \texttt{CV-21} and 8.85\% WER on the \texttt{FLEURS-RO} dataset. Lastly, on the underrepresented speech dataset \texttt{USPDATRO}, we report a 16.5\% relative WER reduction, underscoring the potential for further advancement in Romanian ASR, particularly for low-resource or domain-specific conditions. In terms of inference speed, this method exhibits a $16\%$ relative reduction compared to the CTC greedy decoding strategy. However, it achieves an improvement of over $64\%$ relative to the TDT-ALSD approach, while consistently delivering significantly higher transcription quality across all evaluation datasets.



\section{Conclusions}

In this work, we introduce a state-of-the-art Automatic Speech Recognition (ASR) system for Romanian, leveraging the FastConformer architecture for the first time in this context. By combining over 2600 hours of manually and weakly labeled Romanian speech data, we demonstrate that modern end-to-end architectures—when properly adapted and fine-tuned—can significantly surpass existing systems, including both traditional hybrid HMM-DNN models and large multilingual transformers like Whisper.

Our exhaustive evaluation across seven diverse Romanian benchmarks—including read, spontaneous, oratory, and dialectal speech—confirms the robustness of our system compared to other evaluated systems. We report consistent Word Error Rate (WER) improvements across all test sets, establishing new state-of-the-art results on each. Furthermore, our exploration of multiple decoding strategies, including CTC beam search with a 6-gram token-level language model and TDT-based decoding with ALSD, provides valuable insights into the trade-offs between transcription accuracy and computational efficiency.

To support future research and reproducibility in Romanian speech processing, we commit to publicly releasing our trained model, along with complete training and inference recipes, as well as standardized evaluation datasets. We believe this open-source contribution will help accelerate progress in the broader field of low-resource ASR and foster more inclusive, language-diverse speech technologies.

\textbf{Acknowledgements.} This work was co-funded by EU Horizon project AI4TRUST (No. 101070190) and by a grant of the Ministry of Research, Innovation and Digitization, CNCS/CCCDI - UEFISCDI, project number PN-IV-P8-8.1-PRE-HE-ORG-2023-0078, within PNCDI IV.

\bibliographystyle{IEEEtran}
\bibliography{refs}
\end{document}